\documentclass[aps,twocolumn,prl,superscriptaddress,tightenlines,longbibliography]{revtex4-1}
\usepackage{CJK}
\usepackage{amsfonts}
\usepackage{graphicx}
\usepackage{hyperref}
\usepackage{amsmath}
\usepackage{amssymb}
\usepackage[usenames, dvipsnames]{color}
\usepackage{subfigure}
\usepackage{lipsum}
\usepackage{lineno}

\begin{document}

\title{Coherent Control of an Embedded Bound State Without a Spectral Gap}
\author{Yue Chang}
\email{yuechang7@gmail.com}
\affiliation{Beijing Academy of Quantum Information Sciences, Beijing
100193, China}

\begin{abstract}
Bound states in the continuum (BICs) can confine photonic excitations in
open systems without conventional cavities or band gaps, making them
promising candidates for long-lived quantum storage and single-photon
control. Their use as quantum resources is limited, however, by two basic
difficulties: they are dark to incident photons, and they are not protected
by a spectral gap from the surrounding continuum. We address these
difficulties in a giant atom coupled to a one-dimensional waveguide by
separating radiative access from BIC-preserving manipulation.
Atomic-frequency modulation dynamically breaks and restores the
destructive-interference condition, allowing a mode-matched single photon to
be captured into, and released from, the BIC. Symmetric coupling modulation,
in contrast, preserves the BIC condition throughout the evolution and
continuously changes the atomic and photonic weights of the stored state.
This BIC-preserving deformation realizes a gapless adiabatic control
problem: for a regular embedded continuum, the intrinsic leakage probability
is proportional to the ramp rate rather than quadratically suppressed as in
gapped systems. The linear scaling originates from near-resonant scattering
states at the BIC energy and is independent of microscopic dispersion except
through the prefactor. These results establish a coherent
capture--control--release protocol for a populated BIC and identify the
fundamental error law for manipulating quantum states that remain embedded
in a gapless environment.
\end{abstract}

\maketitle

\textit{Introduction---} Bound states in the continuum (BICs) are localized
eigenstates with energies inside a continuum of propagating modes. Their
radiative decay is suppressed by symmetry, destructive interference, or
parameter tuning. Since the construction of von Neumann and Wigner~\cite%
{vonNeumann1929}, BICs have become a general wave phenomenon in photonic,
acoustic, electronic, and matter-wave systems~\cite{Hsu2016,Sadreev2021}. In
photonics, they enable light confinement in open structures, leading to
ultrahigh-$Q$ resonances, enhanced near fields, and narrow spectral features
with applications in lasing, nonlinear optics, sensing, wavefront control,
and topological photonics~\cite%
{Hsu2013,Zhen2014,Yoon2015,Koshelev2018,Kang2023, Kodigala2017,Luo2024}.

For quantum applications, however, the same interference that protects a BIC
also creates a control problem. An ideal BIC is dark to the continuum and
therefore cannot be loaded by an incoming scattering state. Previous work
has shown that initially excited emitters can relax into stationary
emitter--photon bound states~\cite{Tufarelli2013,Facchi2016}, BICs can be
accessed through nonlinear multiphoton scattering~\cite{Calajo2019}, and
related giant-atom or waveguide-QED settings can realize catch--release
protocols, phase-controlled BICs, and BIC-mediated state transfer~\cite%
{XuGuo2024,ZhangWang2024,GuoWang2025}. These results establish ways of
accessing or engineering embedded bound states. They do not address whether
a populated BIC can be coherently deformed while it remains embedded in a
gapless continuum, nor what error law governs such a process.

This question is not a standard adiabatic-control problem. In a gapped
system, finite-time errors are suppressed by finite energy separation from
other eigenstates. For a BIC, the target state remains resonant with a
continuum of scattering states, so no such gap exists. The issue is whether
destructive interference can protect the instantaneous eigenstate during
deformation, and how near-resonant continuum modes limit the fidelity. This
is central to using BICs as controllable quantum states, since loading,
local manipulation, storage, and readout may favor different atomic and
photonic weights.

Giant atoms in waveguide quantum electrodynamics (QED) provide a natural
setting for this problem. Waveguide QED enables strong and controllable
single-photon light--matter interactions in one-dimensional continua~\cite%
{ShenFan2007,Shi2015,Roy2017,Gu2017,Sheremet2023,Chang2016PRL,
Yang2024PRA,Zanner2022,Facchi2016,Shi2016,Shi2018PRL}, while giant atoms add
spatially separated coupling points whose emitted fields acquire propagation
phases and interfere~\cite%
{Kockum2014,Gustafsson2014,Guo2017,Kockum2018,Kannan2020,Kockum2021}. This
nonlocal interference gives frequency-dependent decay rates, Lamb shifts,
decoherence-free interactions, and non-Markovian multiphoton dynamics~\cite%
{Kockum2014,Kockum2018,Kockum2021,Chang2025CommunPhys}. For special
propagation phases, destructive interference between radiative channels
produces a dressed BIC~\cite%
{Tufarelli2013,GonzalezBallestero2013,Calajo2019,Wang2021}.

\begin{figure*}[tbp]
\centering
\includegraphics[width=\linewidth]{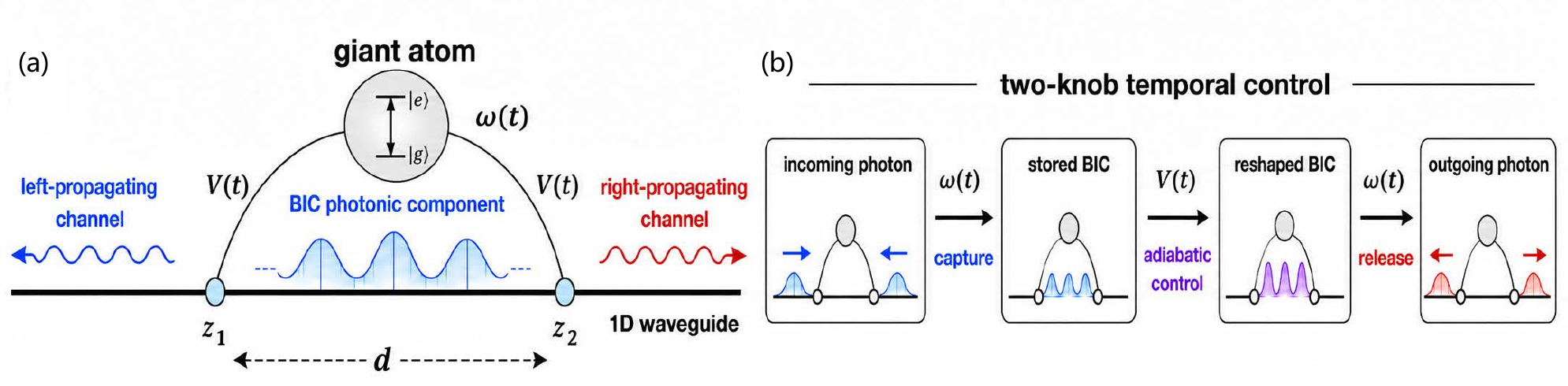}
\caption{Two-knob temporal control of a giant-atom BIC. (a) A two-level
giant atom couples to a one-dimensional waveguide at two spatially separated
points $z_{1}$ and $z_{2}$, with separation $d=z_{2}-z_{1}$. For the BIC,
the photonic component is confined between the two coupling points. The
coupling strength $V(t)$ and the atomic frequency $\protect\omega (t)$
provide independent control knobs. (b) Control protocol. Detuning modulation
$\protect\omega (t)$ dynamically breaks and restores the
destructive-interference condition, enabling capture of a mode-matched
incoming single-photon wavepacket into the BIC and release into an outgoing
photon. Coupling modulation $V(t)$, in contrast, preserves the BIC condition
and adiabatically reshapes the stored embedded state by changing its atomic
and photonic weights.}
\label{fig1}
\end{figure*}

In this Letter, we study a minimal two-point giant atom coupled to a
bidirectional one-dimensional waveguide, as shown in Fig.~\ref{fig1}(a). The
system provides two independent controls, illustrated in Fig.~\ref{fig1}(b).
Atomic-frequency modulation breaks and restores the destructive-interference
condition, providing radiative access to the BIC and enabling capture and
release of a mode-matched single photon. Symmetric coupling modulation, in
contrast, preserves the BIC condition and changes the atomic and photonic
weights of the stored state without intentionally opening a decay channel.
We show that this second operation realizes adiabatic control without a
spectral gap. For a regular embedded continuum, the intrinsic leakage
probability is linear in the ramp rate, rather than quadratically suppressed
as in gapped systems. The dispersion, density of states, and giant-atom form
factor enter only through the prefactor. We verify this behavior for
different ramp shapes and for both linear and cosine dispersions, and
include the effect of realistic imperfections.

\textit{Model and BIC---} The model setup is shown in Fig.~\ref{fig1}.
Setting the photon velocity to unity, the Hamiltonian is
\begin{eqnarray}
H &=&\omega \sigma ^{\dag }\sigma +\int dz\left( i\psi _{L}^{\dag }\partial
_{z}\psi _{L}-i\psi _{R}^{\dag }\partial _{z}\psi _{R}\right)  \notag \\
&&+\sum_{j=1,2}V_{j}\sigma \left[ \psi _{L}^{\dag }\left( z_{j}\right) +\psi
_{R}^{\dag }\left( z_{j}\right) \right] +H.c.
\end{eqnarray}%
Here, $\sigma =|g\rangle \langle e|$, $\omega$ is the atomic transition
frequency, and $\psi _{L(R)}\left( z\right) $ is the annihilation operator
of the left (right)-propagating waveguide field at point $z$, $V_{j}$ is the
atom-waveguide coupling constant at the position $z_{j}$. In the
single-excitation sector, the BIC condition is $V_{1}=V_{2}$ and $\omega
d=(2n+1)\pi $, or $V_{1}=-V_{2}$ and $\omega d=2n\pi $, where $n$ is an
integer. In the following, we focus on the symmetric case $V_{1}=V_{2}=V$.
The energy of the BIC is $\omega $ and the state is
\begin{equation}
|B\rangle =\frac{1}{\sqrt{\mathcal{N}}}\left[ \sigma ^{\dagger
}-iV\int_{0}^{d}dz\sum_{\alpha =L,R}m_{\alpha }e^{im_{\alpha }\omega z}\psi
_{\alpha }^{\dagger }(z)\right] |0\rangle ,  \label{3}
\end{equation}%
where $|0\rangle $ is the ground state with no atomic or photonic
excitation, $m_{R}=-m_{L}=1$, the normalization factor $\mathcal{N}%
=1+2V^{2}d $. The BIC therefore contains an atomic component and a photonic
component confined between the two coupling points.

Under temporal modulation of $\omega (t)$ and $V(t)$, a general
single-excitation state can be written as%
\begin{equation}
\left\vert \psi \left( t\right) \right\rangle =c_{e}\left( t\right) \sigma
^{\dag }\left\vert 0\right\rangle +\int dz\sum_{\alpha =L,R}u_{\alpha
}\left( z,t\right) \psi _{\alpha }^{\dag }\left( z\right) |0\rangle ,
\end{equation}%
where $c_{e}\left( t\right) $ is the atomic excitation amplitude, $u_{\alpha
}\left( z,t\right) $ is the wavefunction of the left- or right-propagating
photon at time $t$ and position $z$. For the symmetric geometry considered
here, the field amplitudes satisfy $u_{L}\left( z,t\right) =u_{R}\left(
d-z,t\right) $ \cite{SupplementalMaterial}.

\begin{figure}[tbp]
\centering
\includegraphics[width=\linewidth]{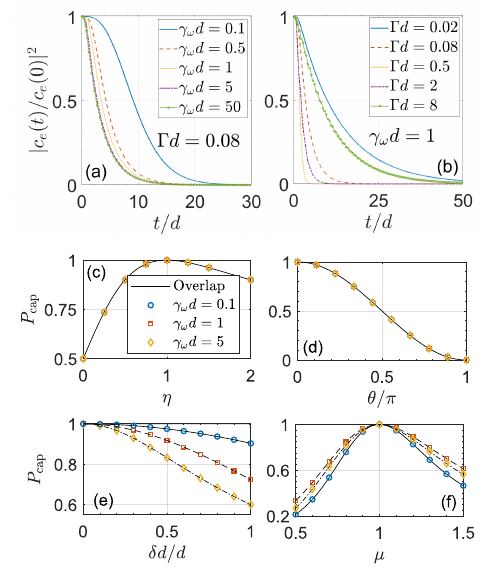}
\caption{Release and capture of the BIC. (a) Decay of the normalized atomic
population $|c_e(t)/c_e(0)|^2$ for different detuning ramp rates $\protect%
\gamma_\protect\omega$, with $\Gamma d=0.08$. (b) Same quantity for
different couplings $\Gamma$, with $\protect\gamma_\protect\omega d=1$.
(c)--(f) Capture probability under input-mode mismatch for $\Gamma d=2$ and
several $\protect\gamma_\protect\omega$. The solid, dashed, and dash-dotted
lines show the mode-overlap prediction $|\langle\Psi_{\mathrm{opt}}|\Psi_{%
\mathrm{in}}\rangle|^2$ for $\protect\gamma_\protect\omega d=0.1$, $1$, and $%
5$, respectively.}
\label{fig2}
\end{figure}

\textit{Capture and release---} The first control knob is the atomic
detuning. We keep the coupling $V$ fixed and write $\omega (t)=\omega
_{0}+\Delta (t)$, where $\omega _{0}d=(2n+1)\pi $ satisfies the BIC
condition for the symmetric coupling configuration. For release, the system
is initialized in the BIC at $\Delta (0)=0$, and $\Delta (t)d\neq 2m\pi $
for any integer $m$, is then turned on. In the rotating frame with respect
to the frequency $\omega _{0}$, $c_{e}(t)=\tilde{c}_{e}(t)e^{-i\omega _{0}t}$%
, the exact single-excitation dynamics reduces to the delay equation
\begin{equation}
\partial _{t}\tilde{c}_{e}(t)=-i\Delta (t)\tilde{c}_{e}(t)-2V^{2}\tilde{c}%
_{e}(t)+2V^{2}\tilde{c}_{e}(t-d),  \label{1}
\end{equation}%
with the BIC history $\tilde{c}_{e}(t\leq 0)=1/\sqrt{\mathcal{N}}$. The last
term in Eq.~(\ref{1}) describes the field emitted at one coupling point and
reabsorbed at the other after the propagation time $d$. Eq.~(\ref{1}) shows
explicitly that detuning does not simply change the phase of the stored
excitation: it breaks the delayed destructive interference and converts the
BIC into a radiating state. Taking an exponential detuning protocol $\Delta
(t)=\Delta _{0}\left( 1-e^{-\gamma _{\omega }t}\right) $, we plot the
normalized atomic population $\left\vert c_{e}\left( t\right)
/c_{e}(0)\right\vert ^{2}$ for different ramp rate $\gamma _{\omega }$ in
Fig.~\ref{fig2}(a). As $\gamma _{\omega }$ increases, the system reaches the
radiative detuned regime more rapidly, leading to a faster decay of the
atomic population. In contrast, the decay is not a monotonic function of the
atom-waveguide coupling strength $\Gamma =2V^{2}$, as shown in Fig.~\ref%
{fig2}(b). As $\Gamma $ increases, the decay first becomes faster and then
slows down. The reason is that increasing $\Gamma $ has two competing
effects: it enhances the bare radiative coupling, but it also increases the
delayed feedback and makes the BIC more photonic. At large $\Gamma $, the
state is mostly stored in the photonic component between the coupling
points, and delayed feedback dominates over local emission. The trapping
effect then wins, and the long-time release becomes slower.

If a detuning protocol $\Delta (t)$ releases the BIC into an outgoing
wavepacket, the time-reversed detuning profile captures the time-reversed
incident state $\left\vert \Psi _{opt}\right\rangle $ back into the BIC.
This optimal mode-matched incident state consists of a pair of
counterpropagating wavepackets satisfying the same spatial symmetry, and can
in principle be generated on demand using a driven three-level emitter in a
cavity- or waveguide-QED interface, where the temporal envelope is set by
the classical control field~\cite%
{Cirac1997,Keller2004,Dilley2012,Shen2023,Stobinska2009}.

Mode mismatch reduces the capture probability $P_{\mathrm{cap}}$ \cite%
{SupplementalMaterial}. We consider four sources of mismatch: a relative
phase $\theta$ and amplitude imbalance $\eta$ between the two input arms, a
timing error $\delta d$, and a pulse-shape mismatch $\mu$. In Figs.~\ref%
{fig2}(c)--(f), we plot $P_{\mathrm{cap}}$ as a function of these mismatches
for $\Gamma d=2$ and $\gamma_\omega d=0.1,1,5$, with an evolution time $40d$%
. Figures~\ref{fig2}(c) and (d) show that $P_{\mathrm{cap}}$ is independent
of the detuning-modulation rate $\gamma_\omega$ for phase and amplitude
imbalance. This follows from the long-time relation $P_{\mathrm{cap}%
}=|\langle \Psi_{\mathrm{opt}}|\Psi_{\mathrm{in}}\rangle|^2$. When only
phase and amplitude imbalance are present, one obtains $P_{\mathrm{cap}}=
\frac{1+\eta^2+2\eta\cos\theta}{2(1+\eta^2)} $. Thus a single-sided input,
corresponding to $\eta=0$, gives $P_{\mathrm{cap}}=1/2$.

Timing errors behave differently. Faster detuning modulation leads to a
stronger reduction of $P_{\mathrm{cap}}$ with increasing $\delta d$, because
the earlier-arriving wavepacket leaks more strongly before destructive
interference is established. For pulse-shape mismatch, the overlap between
the optimal wavepacket $\tilde u_R(z,+\infty)$ and the rescaled wavepacket $%
\tilde u_R(\mu z,+\infty)$ is generally nonmonotonic in the pulse width.
Consequently, for a given mismatch, the capture probability can be optimized
by choosing an appropriate modulation rate $\gamma_\omega$.

\textit{Adiabatic control---} The detuning protocol therefore solves the
radiative-access problem, but it does so by temporarily breaking the BIC
condition. We now turn to the central control problem: reshaping a populated
BIC while the destructive-interference condition remains satisfied
throughout the evolution. We keep the atomic frequency fixed at the BIC
value $\omega (t)=\omega _{0}$ and modulate the symmetric coupling strength $%
V\left( t\right) $. The destructive-interference condition is then preserved
at all times. Therefore, the instantaneous Hamiltonian always has a BIC $%
|B\left( t\right) \rangle $ as its instantaneous eigenstate, with the same
form as Eq.~(\ref{3}), but with time-dependent coupling amplitude $V\left(
t\right) $ and normalization factor $\mathcal{N}\left( t\right)
=1+2V^{2}\left( t\right) d$. Changing $V\left( t\right) $ therefore reshapes
the stored BIC without intentionally opening the radiative channel. If the
system follows the instantaneous BIC, the atomic and photonic weights are $%
P_{e}\left( t\right) =1/\left[ 1+2V^{2}\left( t\right) d\right] $ and $P_{%
\mathrm{ph}}\left( t\right) =2V^{2}\left( t\right) d/\left[ 1+2V^{2}\left(
t\right) d\right] $, respectively. Thus coupling modulation continuously
tunes the BIC between more atom-like and more photon-like configurations:
the former is more locally addressable, whereas the latter is better
protected from emitter decay or dephasing.

The key point is that this control is carried out without a spectral gap.
Although the instantaneous BIC remains an eigenstate throughout the
protocol, its energy is embedded in the scattering continuum. The usual
adiabatic argument based on a finite separation from other eigenstates
therefore does not apply. To quantify the resulting deformation, we study
the full-time dynamics and expand the state in the instantaneous BIC and
scattering eigenstates,
\begin{equation}
|\psi (t)\rangle =b(t)|B\left( V\right) \rangle +\sum_{\alpha =\pm }\int
dk\,A_{\alpha k}(t)|\psi _{k}^{\alpha }\left( V\right) \rangle ,
\end{equation}%
where $|\psi _{k}^{\alpha }\left( V\right) \rangle $ is the normalized
single-photon scattering state with incident momentum $k$, and $\alpha =\pm $
labels the states with right- and left-propagating incident photons,
respectively. Treating the nonadiabatic coupling from $|B(t)\rangle $ to the
scattering continuum perturbatively, the leading scattering amplitude is
\begin{equation}
A_{\alpha k}^{(1)}(t)=\int_{0}^{t}dt^{\prime }e^{-ik(t-t^{\prime
})}e^{-i\omega _{0}t^{\prime }}\dot{V}(t^{\prime })M_{k}^{\alpha }\left[
V\left( t^{\prime }\right) \right] .
\end{equation}%
where the transition amplitude $M_{k}^{\alpha }(V)\equiv -\langle \psi
_{k}^{\alpha }\left( V\right) |\partial _{V}B\left( V\right) \rangle $ is
finite as $k\rightarrow \omega _{0}$. This makes controlled adiabatic
deformation possible despite the absence of a gap.

The adiabatic control error is quantified by the leakage probability $P_{%
\mathrm{leak}}=\sum_{\alpha =\pm }\int dk\,\left\vert A_{\alpha
k}(t)\right\vert ^{2}$ into the scattering channel. For a slow ramp, the
dominant contribution comes from near-resonant scattering states and $P_{%
\mathrm{leak}}$ can be approximated as%
\begin{equation}
P_{\mathrm{leak}}\simeq 2d^{2}\int dt\,\frac{\dot{V}^{2}(t)}{\left[
1+2V^{2}(t)d\right] ^{3}}.  \label{2}
\end{equation}%
Thus, for a ramp of duration $T$, where $\dot V(t)\sim 1/T$, Eq.~(\ref{2})
implies $P_{\mathrm{leak}}\propto 1/T$.

\begin{figure}[tbp]
\centering
\includegraphics[width=\linewidth]{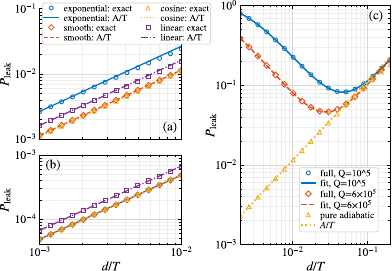}
\caption{Adiabatic control of the BIC. (a) Leakage probability $P_{\mathrm{%
leak}}$ versus ramp rate $1/T$ for four ramp shapes, with $r(0)=0.1$ and $%
r(T)=10$. (b) Same as (a), but with $r(0)=4$ and $r(T)=1$. The markers
denote the numerical results, while the lines show the corresponding
theoretical predictions $P_{\mathrm{leak}}=A/T$ from Eq.~\protect\ref{2}.
(c) Leakage for a smooth ramp including experimental imperfections. The pure
adiabatic leakage and its linear approximation are shown for comparison.}
\label{fig3}
\end{figure}

Since the atomic and photonic populations depend only on $2V^2d$, we define $%
r(t)\equiv 2V^2(t)d=r(0)+\Delta r\, f(s), \qquad s=t/T$, where $T$ is the
ramp duration and $s\in[0,1]$. To test the dependence on the ramp shape, we
consider four choices: $f(s)=\frac{1-e^{-s}}{1-e^{-1}}$, f(s)=s$,
f(s)=3s^2-2s^3$, $f(s)=\frac{1-\cos(\pi s)}{2}$, corresponding to
exponential, linear, smooth, and cosine ramps, respectively. Using the exact
delay-differential dynamics for $\tilde c_e(t)$, together with the field
wavefunction $\tilde u_R(z,t)$ determined by $\tilde c_e(t)$, we obtain the
leakage $P_{\mathrm{leak}}(T)$ shown in Fig.~\ref{fig3}(a). Here $r(0)=0.1$
and $r(T)=10$, so the BIC is changed from an atom-like state to a more
photon-like state. For $T$ ranging from $10^2d$ to $10^3d$, the analytical
prediction of Eq.~(\ref{2}) agrees with the exact numerical result for all
four ramp shapes. A similar agreement is obtained for the reverse-type ramp $%
r(0)=4$ and $r(T)=1$, shown in Fig.~\ref{fig3}(b). In both cases the leakage
is linear in the ramp rate, while the prefactor $A$ depends on the ramp
shape. In particular, because the integrand in Eq.~(\ref{2}) is weighted
more strongly when $2V^2(t)d$ is small, smooth and cosine ramps with reduced
endpoint velocity are advantageous when the ramp starts from, or ends near,
an atom-like state.

Experimental imperfections can add errors that grow with the ramp time and
therefore modify the pure adiabatic scaling. For the linear-dispersion
estimate \cite%
{Luschmann2023ThinFilmSAW,tuokkola2025methods,Gustafsson2014PropagatingPhonons,Andersson2019Nonexponential,Manenti2016SAWResonators,Kongbrailatpam2025FerroelectricHBAR}%
, we take a qubit frequency around $2.95\,\mathrm{GHz}$ and choose the delay
time $d=10\,\mathrm{ns}$. We use representative high-coherence transmon
values $T_{1}=425\,\mu \mathrm{s}$ and $T_{2}^{\mathrm{echo}}=541\,\mu
\mathrm{s}$, a detuning error of $1\,\mathrm{MHz}$, and a coupling imbalance
of $0.5\%$. We compare an acoustic quality factor $Q=10^{5}$ with an
optimistic high-$Q$ case $Q=6\times 10^{5}$ in Fig.~\ref{fig3}(c), with a
smooth ramp from $r(0)=0.1$ to $r(T)=10$. The total leakage is well
described by $1-\exp (-A/T-BT)$, where the term $A/T$ is the intrinsic
adiabatic leakage and $BT$ accounts for the imperfections \cite%
{SupplementalMaterial}. The optimal ramp rate is therefore $1/T_{\mathrm{opt}%
}=\sqrt{B/A}$. For $Q=10^{5}$ and $Q=6\times 10^{5}$, the minimum leakage is
approximately $0.08$ and $0.05$, respectively.

\textit{Universality of the linear scaling---} Equation~(\ref{2}) is the
linear-spectrum form of a more general regular-continuum result. Let $%
|B(V)\rangle $ be an instantaneous BIC with energy $\omega _{B}$ embedded in
scattering states $|\psi _{k}^{\alpha }(V)\rangle $ with dispersion $\omega
_{k}$. The relevant adiabatic spectral density is
\begin{equation}
\mathcal{J}_{\mathrm{ad}}(\omega ,V)=\sum_{\alpha }\int dk\,\left\vert
M_{k}^{\alpha }(V)\right\vert ^{2}\delta (\omega -\omega _{k}).
\label{eq:jad}
\end{equation}%
This quantity includes the local density of states, or equivalently the
inverse group velocity for a dispersive waveguide. If $\mathcal{J}_{\mathrm{%
ad}}(\omega ,V)$ is smooth on the near-resonant window $|\omega -\omega
_{B}|\lesssim 1/T$ which is well-located in the band, the leading leakage is
\begin{eqnarray}
P_{\mathrm{leak}} &\simeq &2\pi \int dt\,\dot{V}^{2}(t)\mathcal{J}_{\mathrm{%
ad}}(\omega _{B},V(t))  \notag \\
&=&\frac{2\pi }{T}\int_{0}^{1}ds\,\left[ \partial _{s}V(s)\right] ^{2}%
\mathcal{J}_{\mathrm{ad}}(\omega _{B},V(s))  \label{4}
\end{eqnarray}%
for a finite ramp. Thus the linear scaling $P_{\mathrm{leak}}\propto 1/T$ is
universal for a regular embedded continuum. The ramp shape, waveguide
dispersion, local group velocity, and giant-atom form factor enter the
prefactor.

\begin{figure}[tbp]
\centering
\includegraphics[width=\linewidth]{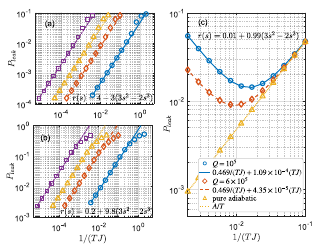}
\caption{Adiabatic control in a cosine-dispersive waveguide. (a),(b) Leakage
probability $P_{\mathrm{leak}}$ versus ramp rate $1/T$ for smooth ramps.
From right to left, the circles, diamonds, triangles, and squares denote the
numerical results for $D=2$, $10$, $20$, and $50$, respectively. The solid,
dashed, dotted, and dash-dotted lines show the corresponding theoretical
predictions $P_{\mathrm{leak}}=A/T$ from Eq.~\protect\ref{4}. (c) Leakage
for a smooth ramp including system imperfections. The pure adiabatic leakage
and its linear approximation are shown for comparison. }
\label{fig4}
\end{figure}

This linear scaling differs from the usual finite-time gapped adiabatic
result, where the leakage probability is generically quadratic in the ramp
rate. In a gapped system, every leakage channel has a finite energy mismatch
$\Delta _{n}$, so the phase factor $\exp \left[ i\int^{t}\Delta
_{n}(t^{\prime })dt^{\prime }\right] $ oscillates throughout the protocol.
Integration by parts then gives a transition amplitude $A_{n}\sim 1/T$ and
hence a probability $P_{n}\sim 1/T^{2}$. For the BIC, however, the leakage
states form a continuum with detuning $\Delta _{k}=\omega _{k}-\omega _{B}$
that can be arbitrarily small. Modes with $\left\vert \Delta _{k}\right\vert
\lesssim 1/T$ do not acquire enough phase to average out during the ramp.
The transition amplitude into each such near-resonant mode is set by the
total change of the BIC and is not suppressed by an additional gap factor,
while the width of the resonant window is $\sim 1/T$. Integrating over these
modes therefore gives $P_{\mathrm{leak}}\sim 1/T$. Equivalently, the
continuum converts the squared ramp velocity into an instantaneous leakage
rate, $\Gamma _{\mathrm{leak}}\left( t\right) \sim \dot{V}\left( t\right)
^{2}$, and the total probability scales as $\int_{0}^{T}dt\,\dot{V}\left(
t\right) ^{2}\sim 1/T$.

As an illustration, Figs.~\ref{fig4}(a) and (b) show the linear scaling for
a cosine band \cite{Megrant2012APL,Scigliuzzo2022PRX,Jouanny2025NatCommun}
using a smooth ramp. Here $D$ is the separation between the two coupling
points and $4J$ is the bandwidth. We define $r(s)=2V^2(s)D/v_{k_B}^2$, where
$k_B$ is the momentum corresponding to the BIC energy $\omega_B$, and $%
v_k=2J|\sin k|$ is the group velocity. In contrast to the linear-dispersion
case, the ramp time required to reach the asymptotic linear-scaling regime
depends on the BIC energy, because the density of states is proportional to $%
1/v_k$. Close to a band edge, the maximum ramp rate compatible with the
asymptotic $1/T$ regime scales as $D^{-2}$.

We also consider the total leakage in the presence of experimental
imperfections. For $D=2$, corresponding to a BIC in the middle of the band,
we take an atomic frequency of $5\,\mathrm{GHz}$, $J=2\pi\times150\,\mathrm{%
MHz}$, and the same imperfection parameters as in Fig.~\ref{fig3}(c). The
resulting leakage for two quality factors is shown in Fig.~\ref{fig4}(c). In
this regime, $P_{\mathrm{leak}}(T)$ is well described by $A/T+BT$ \cite%
{SupplementalMaterial}, reflecting the competition between pure adiabatic
leakage and imperfections. The corresponding optimal ramp times are
approximately $70\,\mathrm{ns}$ and $110\,\mathrm{ns}$ for $Q=10^5$ and $%
Q=6\times10^5$, respectively. The imperfection-induced contribution is
smaller than in the linear-dispersion case, mainly because the shorter
separation between the coupling points reduces the effect of finite
waveguide quality factor.

\textit{Conclusion and outlook---} We have studied how a populated
single-excitation BIC can be accessed and then coherently deformed while
remaining embedded in a gapless continuum. The protocol separates radiative
access from BIC-preserving manipulation. Atomic-frequency modulation breaks
the destructive-interference condition and enables capture and release of a
mode-matched photon, whereas symmetric coupling modulation preserves the BIC
condition and changes the atomic and photonic weights of the stored
excitation. The latter operation realizes adiabatic control without a
spectral gap. We derived the leading leakage probability and found that, for
a regular embedded continuum, it is linear in the ramp rate. The dispersion,
density of states, and giant-atom form factor enter only through the
prefactor. This scaling is illustrated for both linear and cosine
dispersions; near a band edge, the onset of the asymptotic regime is shifted
to longer ramp times. Including realistic imperfections gives an additional
time-growing error and therefore an optimal ramp time.

Our results establish a control principle for embedded quantum states: a
stored BIC can be deformed without intentionally opening a radiative
channel, but gapless control imposes an intrinsic linear-in-rate fidelity
limit. This framework can be extended to more complex giant-atom geometries,
multi-emitter networks, and optimized protocols.

\textit{Acknowledgment---} We thank the anonymous referees for comments that
motivated us to examine the generality of the linear scaling. This work was
supported by the National Natural Science Foundation of China under Grant
No.~12575008.


\bibliography{V3.bbl}

\end{document}